\documentclass[conference]{IEEEtran}
\IEEEoverridecommandlockouts
\usepackage{cite}
\usepackage{amsmath,amssymb,amsfonts}
\usepackage{algorithmic}
\usepackage{graphicx}
\usepackage{textcomp}
\usepackage{xcolor}
\usepackage{hyperref}
\def\BibTeX{{\rm B\kern-.05em{\sc i\kern-.025em b}\kern-.08em
    T\kern-.1667em\lower.7ex\hbox{E}\kern-.125emX}}
\begin{document}

\title{SoftPUF: a Software-Based Blockchain Framework using PUF and Machine Learning 
}

\author{\IEEEauthorblockN{S M Mostaq Hossain\textsuperscript{1}, Sheikh Ghafoor\textsuperscript{1}, Kumar Yelamarthi\textsuperscript{2} and Venkata Prasanth Yanambaka\textsuperscript{3}}
\IEEEauthorblockA{\textsuperscript{1}\textit{Dept. of Computer Scinece, Tennessee Technological University
Cookeville, Tennessee, USA} \\
\textsuperscript{2}\textit{College of Engineering, Tennessee Technological University
Cookeville, Tennessee, USA} \\
\textsuperscript{3}\textit{Dept. of Computer Science, Texas Woman's University, Denton, Texas, USA} \\
Email: shossain42@tntech.edu, SGhafoor@tntech.edu, kyelamarthi@tntech.edu and vyanambaka@twu.edu}
}

\maketitle

\begin{abstract}

Physically Unclonable Function (PUF) offers a secure and lightweight alternative to traditional cryptography for authentication due to their unique device fingerprint. However, their dependence on specialized hardware hinders their adoption in diverse applications. This paper proposes a novel blockchain framework that leverages SoftPUF, a software-based approach mimicking PUF. SoftPUF addresses the hardware limitations of traditional PUF, enabling secure and efficient authentication for a broader range of devices within a blockchain network. The framework utilizes a machine learning model trained on PUF data to generate unique, software-based keys for each device. These keys serve as secure identifiers for authentication on the blockchain, eliminating the need for dedicated hardware. This approach facilitates the integration of legacy devices from various domains, including cloud-based solutions, into the blockchain network. Additionally, the framework incorporates well-established defense mechanisms to ensure robust security against various attacks. This combined approach paves the way for secure and scalable authentication in diverse blockchain-based applications. Additionally, to ensure robust security, the system incorporates well-established defense mechanisms against various attacks, including 51\%, phishing, routing, and Sybil attacks, into the blockchain network. This combined approach paves the way for secure and efficient authentication in a wider range of blockchain-based applications.
\end{abstract}


\section{Introduction}
The technological landscape is constantly evolving, with new devices and applications emerging at a rapid pace. 
Traditional cryptographic solutions, while robust, can be complex for resource-constrained devices and often require significant computational power. As this ecosystem grows, so does the need for secure and efficient authentication methods. Traditional cryptographic solutions, while robust, can be complex for resource-constrained devices and often require significant computational power. This complexity creates a challenge: how do we ensure secure authentication for a wider range of devices without sacrificing performance or scalability? Here's where PUFs \cite{bathalapalli2022pufchain} come in. They offer a lightweight and secure alternative by leveraging a device's unique physical characteristics to generate a fingerprint for authentication. Machine learning (ML) models are able to learn the unique challenge-response relationships within a PUF dataset, enabling them to mimic PUF behavior mathematically without physical replication\cite{shamsoshoara2020survey}. However, the current implementation of PUFs relies on specialized hardware, making them unsuitable for a broader range of devices, particularly those lacking the necessary processing power or physical space. 

Blockchain technology, on the other hand, offers a promising avenue for secure and scalable authentication. Blockchain networks function as distributed ledgers, where every transaction is cryptographically recorded and replicated across a network of devices. This distributed ledger ensures the immutability and reliability of data, making it ideal for storing and managing authentication records. Additionally, blockchain protocols facilitate secure communication between devices participating in the network. This inherent security aligns perfectly with the need for robust authentication solutions.

\begin{figure*}
  \centerline{\includegraphics[width=1\linewidth, keepaspectratio]{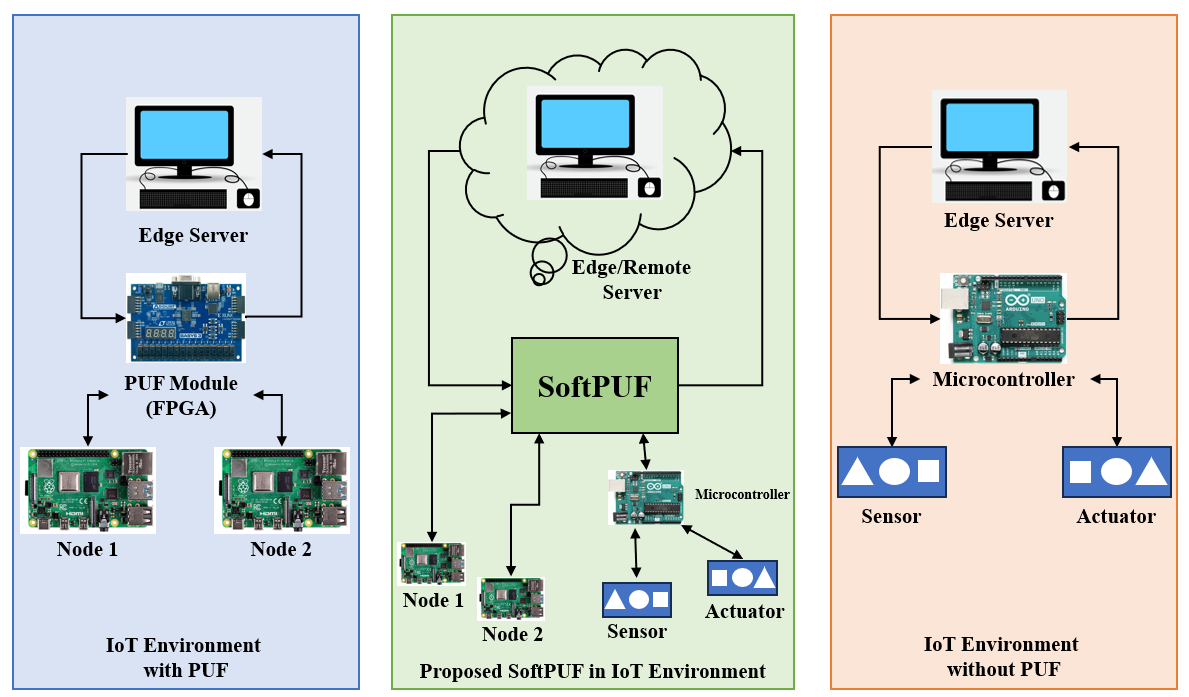}}
  \caption{Overview diagram of SoftPUF.}
  \label{fig}
\end{figure*}




This paper proposes a novel framework, leveraging a software base solution within a blockchain architecture. SoftPUF is a software-based approach that mimics the functionality of PUF. Utilizing machine learning, it leverages training data derived from real PUF challenge-response pairs \cite{lee2020novel} to generate unique keys for each device. These software-generated keys act as secure identifiers, eliminating the need for dedicated hardware and enabling seamless integration with blockchain technology. By addressing the hardware limitations of traditional PUFs, SoftPUF paves the way for secure and efficient authentication in a wider range of blockchain-based applications. This framework unlocks the potential of PUF-like authentication for a broader range of devices, fostering a more secure and trustworthy digital landscape. This research focuses on three key aspects:

Mimicking PUF Characteristics: SoftPUF utilizes a machine learning model trained on PUF data to generate unique and unpredictable keys, replicating the core functionality of hardware-based PUFs.

    Software-based Key Generation: By leveraging machine learning, SoftPUF eliminates the need for dedicated PUF hardware, making it a software-defined solution for secure authentication.

Integration with Blockchain: SoftPUF seamlessly integrates with blockchain technology, leveraging its distributed ledger and secure communication protocols to enhance the overall security and scalability of the authentication process.

By addressing the hardware limitations of traditional PUFs, SoftPUF paves the way for secure and efficient authentication in a wider range of blockchain-based applications. This opens doors for the secure integration of legacy devices and diverse systems into the blockchain ecosystem, fostering a more secure and trustworthy digital landscape. Figure 1 provides an overview of SoftPUF.

\section{Related Prior Research}

Existing PUF-based solutions often rely on specialized hardware, limiting their portability and scalability. This work addresses this challenge by utilizing a machine learning model to emulate PUF behavior, enabling software-based PUF functionality. While blockchain technology has been explored for secure data storage and authentication, its integration with PUFs for generating unique keys and enhancing security has been limited. This work bridges this gap by seamlessly integrating PUF keys into a blockchain network, offering a robust and secure solution. 


In the study \cite{sadhu2022machine}, Sadhu et al. presented a technique that combines PUF and ML to keep data secure and private while also providing a lightweight authentication alternative. PUF is utilized with ML to prevent extra hardware storage space. The authentication between the two messages is completed using only lightweight operations like hash, XOR, concatenation, etc. The computational and communication costs of their system are lower than those of competing solutions. The proposed framework also worked as a combination of PUF and ML with different hyperparameters and better accuracy and was not only dependent upon 64-bit keys but also worked with key combinations. Moreover, Seema et al. proposed a PUF-based certificate authority protocol for edge data center (EDC) authentication in collaborative edge computing, eliminating the need for a resource-intensive CRP database\cite{aarella2023fortified}. While their model tackles secure authentication in edge computing using PUF-based certificates, a key research gap lies in leveraging machine learning for dynamic CRP generation and verification. The paper proposes a static certificate approach, where each EDC stores a fixed set of CRPs. This can be inefficient, especially with dynamic load balancing and a large number of EDCs. Compared to the proposed framework using the dynamic machine-learning model it solved the problem of dynamic CRP generation. Furthermore, Prosanta et al. 
claimed scalability due to the concept of one-time PUF, but it didn't analyze the impact of increased network traffic and reconfiguration overhead on resource-constrained medical IoT devices \cite{gope2021scalable}. Exploring lightweight communication protocols and optimizing the reconfiguration process could improve scalability in real-world settings. And this scope of work is followed in the proposed framework by using a lightweight communication protocol like UDP with a defense technique against well-known attacks. 

Li et al. \cite{li2022blockchain} claimed scalability by having entities authenticate devices instead of storing large CRP databases, but they do not fully address how this is achieved in practice. Scalability challenges in blockchain networks may still arise with an increasing number of IIoT devices and frequent authentication requests. Train an ML model (federated learning could be ideal) on existing CRPs to dynamically generate challenge-response pairs, reducing storage needs on individual devices and scaling efficiently with a growing number of IIoT endpoints. Blockchain would then store only verification keys and metadata, not massive CRP databases. In the paper, Almuqren et al. mainly focused on theoretical analysis and benchmark dataset evaluation \cite{almuqren2023blockchain}. Real-world smart home environments are diverse and often involve resource-constrained devices. While the paper claims scalability due to model-based authentication, the resource overhead of PUF model training and deployment on resource-constrained smart home devices needs further evaluation. The cloud-based structure of the proposed framework solved this issue as it wouldn't have to think about the training and deployment phases overhead. On the other hand, Jia et al. propose a blockchain-assisted authentication framework for Internet of Medical Things systems using fog computing \cite{jia2022blockchain}. It utilizes different authentication protocols based on the entities involved. While secure and efficient, the research acknowledges limitations in further improving efficiency and achieving other security properties. Future work could address these by exploring lighter-weight protocols or incorporating additional security features.

If we talk about Blockchain-based security, Patil et al. proposed a secure authentication protocol for IoT using blockchain and PUFs \cite{patil2020blockchain}. However, the security analysis of the blockchain aspect focuses on the benefits it offers (immutability, tamper-proof records) without mentioning potential vulnerabilities specific to blockchain technology in the context of IoT. Areas for further exploration could include blockchain scalability challenges for many IoT devices or potential security risks associated with smart contracts used in the proposed framework. In another research, Mahmood et al. propose a secure key establishment protocol for cross-domain IIoT using blockchain and PUFs\cite{mahmood2023blockchain}. While the security analysis mentions resistance to specific attacks (desynchronization, impersonation), it doesn't address potential vulnerabilities of the blockchain technology itself in the context of the Industrial Internet of Things (IIoT). Areas for further exploration could be the scalability challenges of blockchain for many IIoT devices or security risks associated with smart contracts used in the proposed framework. Additionally, the research mentions reducing on-chain storage overhead but doesn't elaborate on potential trade-offs between security and efficiency achieved by this approach. However, Felicetti et al. proposed a novel combination of PUF-based tags for unique, unclonable identification and deep learning for detecting reapplied tags on counterfeit items\cite{felicetti2023deep}. However, further research is needed to evaluate the real-world performance of the deep learning models against sophisticated counterfeiting techniques, ensuring it can adapt to ever-evolving attack methods in the complex supply chain landscape.

So, existing PUF-based device security solutions often face scalability and practical challenges as they often rely on specialized hardware, making them difficult to deploy and update across diverse devices. Additionally, storing large databases of CRPs for verification can be inconvenient and expensive, especially for resource-constrained devices. Furthermore, existing approaches often lack dynamic adaptability to evolving threats and sophisticated counterfeiting techniques.


\section{Proposed SoftPUF}
This work proposes SoftPUF, a novel software-based approach to mimic the functionality of PUFs for device authentication. SoftPUF overcomes the hardware limitations of PUFs by leveraging machine learning to generate unique keys for each device based on real PUF data. These software-generated keys eliminate the need for dedicated hardware and enable seamless integration with blockchain technology for secure communication and enhance overall security posture. This research employs the following operations to analyze and validate the effectiveness of the proposed model across its key functionalities.

\subsection{Software-based Unique Key Generation}

This system uses a Machine Learning model to create unique keys that have PUF characteristics. The dataset contains the challenges and responses of PUF. The challenges are inputs for PUF and responses are the corresponding outputs. These challenges and responses create a pair that helps to create the unique keys. According to the dataset nature, a custom Linear Regression model has been proposed. This model eventually mimics the PUF characteristics mathematically and provides the unique keys. These keys are like fingerprints and make sure only authorized devices can connect and communicate securely. 

\begin{figure}[h!]
  \centerline{\includegraphics[width=0.5\textwidth, keepaspectratio]{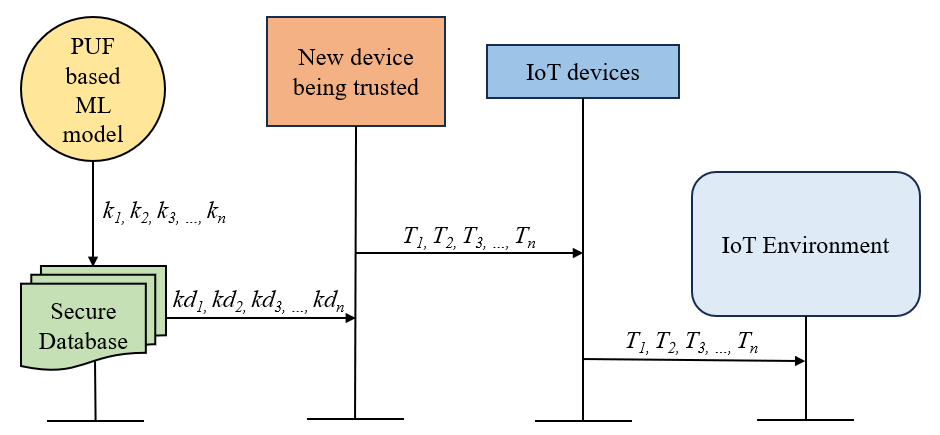}}
  \caption{Device Enrollment Steps}
  \label{fig}
\end{figure}

\subsection{Device Enrollment}
Upon successful generation of the keys, denoted as \textit{K\textsubscript{n}}, by the ML model, a crucial process of device enrollment commences, as shown in Figure 2. This key, a cornerstone of trust for each device, is safeguarded within a secure database as \textit{Kd\textsubscript{n}}, ensuring its integrity and confidentiality throughout the enrollment process. The enrollment process seamlessly integrates with the key generation stage, establishing trusted identities as \textit{T\textsubscript{n}} within the IoT ecosystem. The securely stored keys serve as the foundation for enrolling new devices, enabling their participation as either miners or clients. To facilitate secure communication and authentication, the key \textit{T\textsubscript{n}} is embedded within each trusted device before its deployment into the IoT environment. This type of designed enrollment procedure lays a robust foundation for secure and reliable interactions among devices within the proposed framework.

\subsection{Transaction Initiation}
Devices initiate transactions seamlessly upon network integration, partnering with trusted nodes with key \textit{T\textsubscript{n}} for validation shown in Figure 3. To safeguard identities and intentions, devices encapsulate their MAC addresses \textit{M\textsubscript{n}}, data \textit{k\textsubscript{n}}, and PUF key lengths \textit{N} within a SHA512-encrypted object hash \textit{h\textsubscript{n}}, broadcasting this secure beacon to the network, and anticipating the commencement of trusted interactions.

\begin{figure}[h!]
  \centerline{\includegraphics[width=1\linewidth, keepaspectratio]{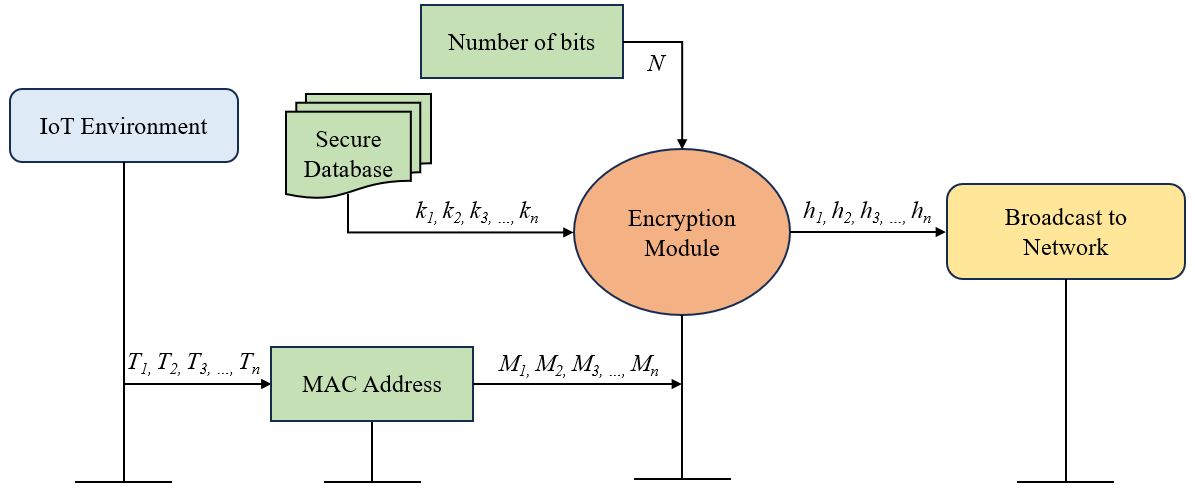}}
  \caption{Initial transaction steps.}
  \label{fig}
\end{figure}

\subsection{Device Authentication}
Prior to inclusion within the blockchain, broadcasted blocks undergo a rigorous authentication process orchestrated by trusted nodes. Upon receiving a block, a trusted node extracts both the object and its corresponding hash. The message is then decrypted, and a pivotal attempt to retrieve the key using the private key ensues. Successful key retrieval signifies a trusted device and prompts secure storage of the key within a dedicated database. Figure 4 shows its details. Conversely, failed retrieval raises alarms, casting doubt upon the device's trustworthiness and barring its entrance into the blockchain, safeguarding the integrity of the network.

\begin{figure}[h!]
  \centerline{\includegraphics[width=1\linewidth, keepaspectratio]{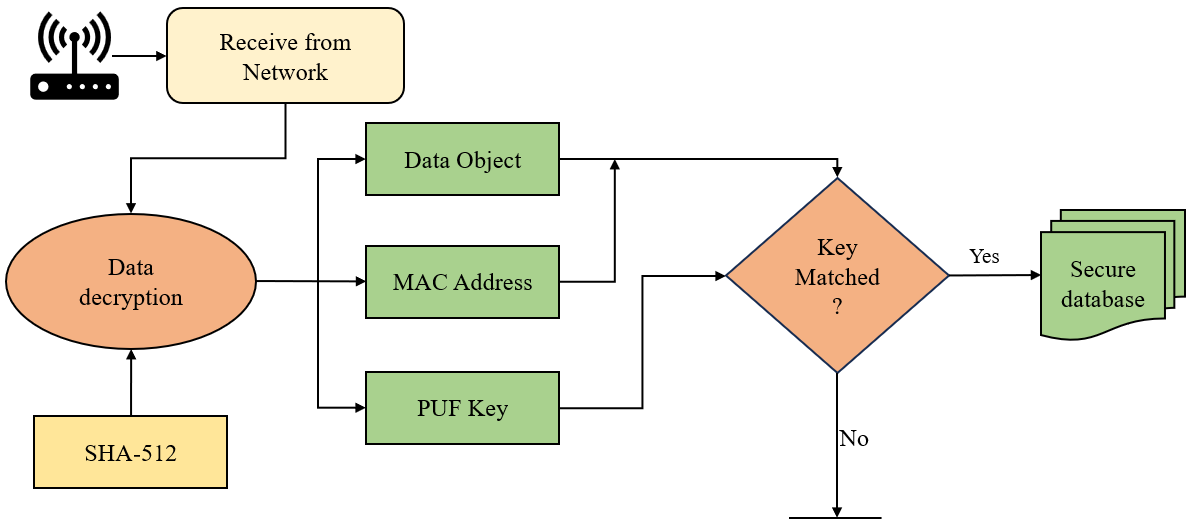}}
  \caption{Device authentication for the trusted nodes.}
  \label{fig}
\end{figure}

\section{Experimental Results}
This section has been divided into five main subsections: experimental setup, design components of the framework, machine learning model specification, result analysis, and defense techniques. Those are the following:
\subsection{Experimental setup}
The experimental design illustrated in Figure 5, commences with the acquisition of PUF datasets (1Mx8) from the PUF module. These datasets embark on a journey to a remotely accessible ML model edge server, hosted within a cloud-based environment constructed using Google Colab Pro. This cloud-powered ecosystem, powered with a 12.7 GB high-RAM runtime, empowers the execution of critical model development stages, encompassing data preprocessing, model generation, key generation, and rigorous evaluation. This architectural decision fosters portability and scalability, laying a robust foundation for future advancements.

\begin{figure}[htbp]
  \centerline{\includegraphics[width=1\linewidth, keepaspectratio]{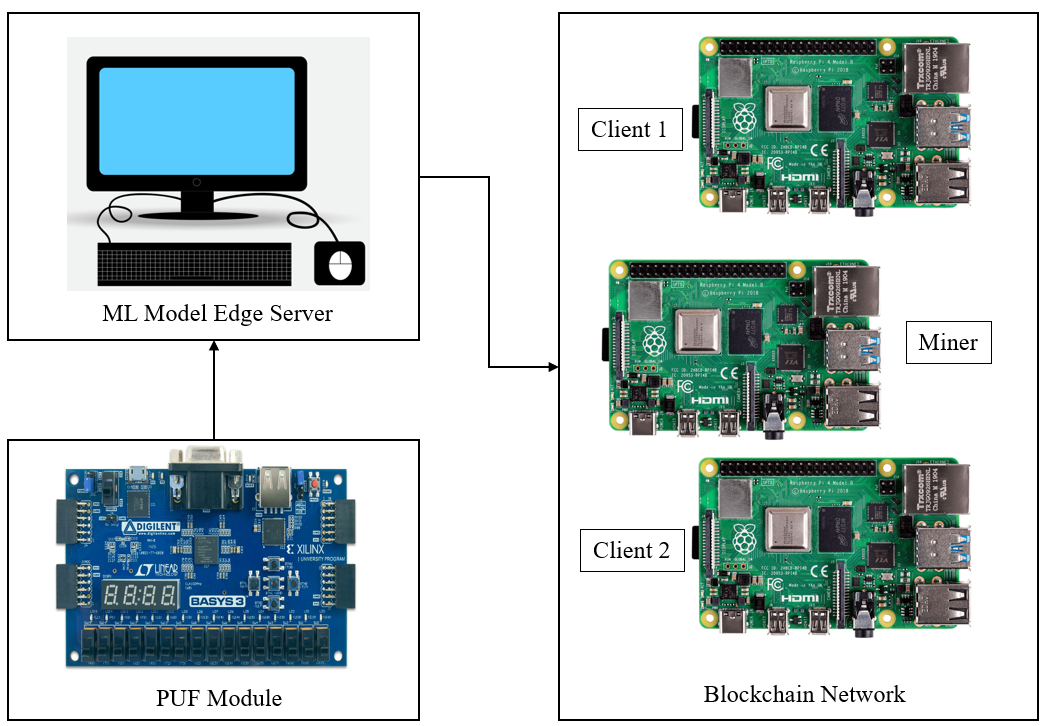}}
  \caption{Experimental Setup of the Testbed.}
  \label{fig}
\end{figure}

\subsection{Framework Design Components}
The culmination of the ML model's execution yields a unique PUF key, a pivotal component that seamlessly integrates with the blockchain network. It is within this network that the key steps are a series of security-centric tasks, encompassing the creation of trusted nodes, authentication procedures, and the facilitation of secure data transactions among diverse clients and miners. This interplay of components culminates in the framework's readiness for application-level deployment, a testament to its robust design and potential for real-world impact. Table II encapsulates a concise overview of the proposed framework's key characteristics.

\begin{table}[]
\centering
\caption{Technical Specifications of Proposed Framework}
\label{tab:my-table}
\begin{tabular}{p{0.45\linewidth}p{0.45\linewidth}}
\hline
\\[-0.8em]
Modules & Components    \\[-0.8em]       \\ \hline
\\[-0.8em]
PUF module                    & 64-bit arbiter PUF   \\ 
PUF and Hashing module        & Xilinx Artix-7 FPGA  \\ 
Edge server                   & Google Colab machine \\ 
Client node                   & Raspberry pi 4       \\ 
Miner node                    & Raspberry pi 4       \\ 
Blockchain database           & SQLite               \\ 
Communication                 & Serial (UART), UDP   \\ 
Programming language          & Python     \\ \\[-0.8em] \hline
\end{tabular}
\end{table}

\subsection{Hyperparameters and Architecture of ML model}
The model's architecture draws upon a combination of hyperparameters, each contributing to its efficacy in classifying data. The \textit{adam} optimizer\cite{mursi2021hybrid}, renowned for its adaptability and efficiency in navigating complex optimization landscapes, steers the learning process. \textit{ReLU} and \textit{tanh} activation functions, positioned within different layers, facilitate both efficient gradient propagation and nuanced feature representation. A kernel initializer \textit{he\textunderscore uniform} establishes the initial weights within the network, ensuring a sound starting point for learning. The \textit{hinge} loss function, chosen for its adeptness in binary classification tasks, guides the model towards clear decision boundaries. \textit{accuracy}, serving as the primary evaluation metric, measures the model's ability to accurately classify data points. Table I summarizes the hyperparameters used to train a deep neural network for classification. The model's architecture features a thoughtfully designed arrangement of neurons across layers, empowering the network to extract intricate features from the input data while mitigating overfitting. Notably, \textit{dropout} and \textit{Z-score} normalization techniques were intentionally left out, suggesting the model was specifically tuned to the specific characteristics of the dataset and the nature of the classification task.

\begin{table}[ht]
\centering
\caption{Model Design Specifications}
\label{tab:my-table}
\begin{tabular}{p{0.45\linewidth} p{0.45\linewidth}}
\hline
\\[-0.8em]
Components & Hyper-parameters   \\[-0.8em]   \\ \hline
\\[-0.8em]
Optimizer & adam \\ 
Units & 512-512-128-128-64-32-8 \\ 
Dropout & × \\ 
Activation & relu, tanh \\ 
Kernel initializer & he\_uniform \\ 
Loss & hinge \\ 
Metrics & accuracy \\ 
Z-Score & × \\ \\[-0.8em] \hline
\end{tabular}
\end{table}

\subsection{Result Analysis}
This framework uses regression to analyze data and make predictions. Two methods are employed to assess the model's efficacy: how good it is at finding connections in the data, and how closely its predictions match real-world values. Different measures, such as R-squared, MSE, MAE, and RMAPE, are employed to gain a clear understanding of the model's strengths and weaknesses. 
These insights, collectively, offer a nuanced understanding of the model's strengths and potential areas for refinement, paving the way for informed decision-making and future research endeavors.

The results encapsulated in Table III unveil a performance comparison of MAE and MSE among the evaluated regression models. \textit{GradientBoostingRegressor} emerges as the frontrunner, boasting the highest R-squared score and the lowest MSE and MAE scores, indicative of a robust fit to the data. \textit{AdaBoostRegressor} follows closely with a commendable R-squared score and a notably low RMAPE score. In contrast, \textit{KNeighborsRegressor} falters with the lowest R-squared score and the highest MSE and MAE scores, suggesting a less optimal alignment with the underlying patterns. The observed variability in performance across different metrics underscores the importance of carefully selecting evaluation criteria tailored to the specific task at hand to ensure the most suitable model is chosen for the intended application.

\begin{table}[]
\centering
\caption{MAE and MSE Comparison for different models}
\label{tab:my-table}
\begin{tabular}{p{0.32\linewidth}p{0.27\linewidth}p{0.26\linewidth}}
\hline
\\[-0.8em]
Model & Mean Absolute Error & Mean Squared Error\\ \\[-0.8em] \hline
\\[-0.8em]
GradientBoostingRegressor   & 0.9927    & 0.9996          \\ 
AdaBoostRegressor           & 0.9930    & 1.0000          \\ 
KNeighborsRegressor         & 0.9927    & 1.1112          \\ 
LinearRegression            & 0.9928    & 0.9998          \\ 
DecisionTreeRegressor       & 1.0405    & 1.5965          \\ 
RandomForestRegressor       & 0.9926    & 0.9995          \\ 
Ridge                       & 0.9928    & 0.9998          \\ 
Proposed ML model    & 0.9946    & 1.6440    \\ \\[-0.8em] \hline
\end{tabular}
\end{table}

While not outperforming the top-tier GradientBoostingRegressor and Ridge models in MSE, the proposed framework model strikes a compelling balance between accuracy and adaptability, offering computational efficiency, interpretability, and potential for tailored applications shown in Figure 6. The MAPE of 1.2768 suggests reasonable, though not exceptional, prediction accuracy, but its competitive MAE of 0.9946 positions it among the frontrunners in terms of low average absolute errors. This model's strengths extend beyond raw performance metrics, making it a promising candidate when computational efficiency, interpretability, or adaptability are of paramount importance.

\begin{figure}[h!]
  \centerline{\includegraphics[width=1\linewidth, keepaspectratio]{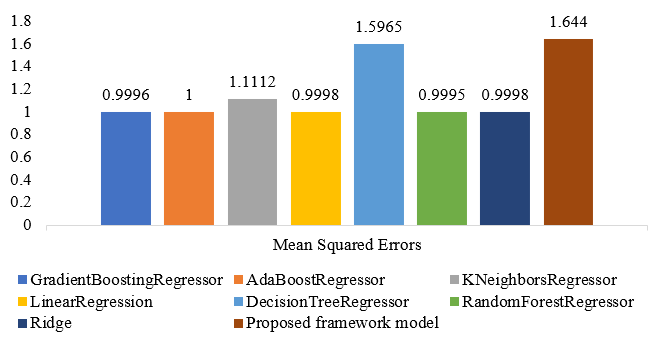}}
  \caption{MSE comparison of different regressor models.}
  \label{fig}
\end{figure}

The key generated by the framework was subsequently submitted to the Blockchain environment. Before doing that the keys have been encrypted with the client’s MAC address and then a SHA512 encryption has been executed to make it secure. 

Figure 7 shows the hash generation process. After the hash generation process, the task for the Blockchain network has been started. From the miner end when it has got the has of the data then it decrypts the data with the hash key and gets it the MAC, Data, and the key length. Then it saves the data to its database to use it further needed.

\begin{table}[]
\centering
\caption{Execution time of different phases in Blockchain network}
\label{tab:my-table}
\begin{tabular}{p{0.45\linewidth}p{0.45\linewidth}}
\hline
\\[-0.8em]
Phase Items & Execution Time (ms) \\ \\[-0.8em] \hline
\\[-0.8em]
Phase 1   & 40.72              \\ 
Phase 2           & 9.83               \\ 
Phase 3         & 0.22             \\ 
Total            & 50.77          \\ \\[-0.8em] \hline
\end{tabular}
\end{table}

An analysis of transaction execution time, conducted using a randomly selected 20\% subset of the 1M dataset, revealed a total processing time of 50.77ms as presented in Table IV. Phase 1, encompassing block generation, consumed 37.8s, predominantly for model prediction. Key generation contributed a negligible 2.77ms. Phase 2, involving block mining and key acceptance within the wireless network, required 9.83ms. Finally, Phase 3, dedicated to block addition to the network, incurred a swift 0.22ms. 

\begin{figure}[h!]
  \centerline{\includegraphics[width=1\linewidth, keepaspectratio]{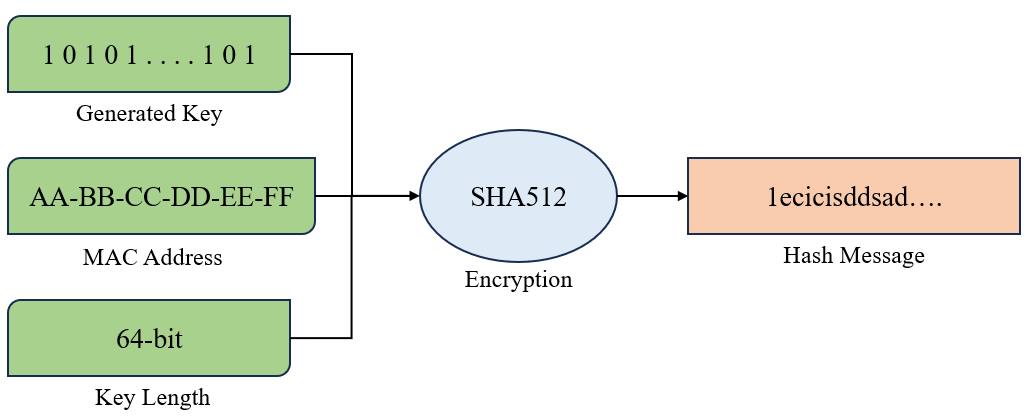}}
  \caption{Hash generation process.}
  \label{fig}
\end{figure}

Figure 8 explains the message flow architecture, and message sizes within each phase. In phase 1, a PUF key is generated from the machine learning model. This key is then carefully combined with a block address and a small piece of information about the key's size in phase 2. This creates a 119-bit message that is then scrambled using a strong encryption method, making it 512 bits long. Phase 3 has a message that's the same size, sent back by the miner to make sure the data is exchanged smoothly.

\begin{figure}[htbp]
  \centerline{\includegraphics[width=1\linewidth, keepaspectratio]{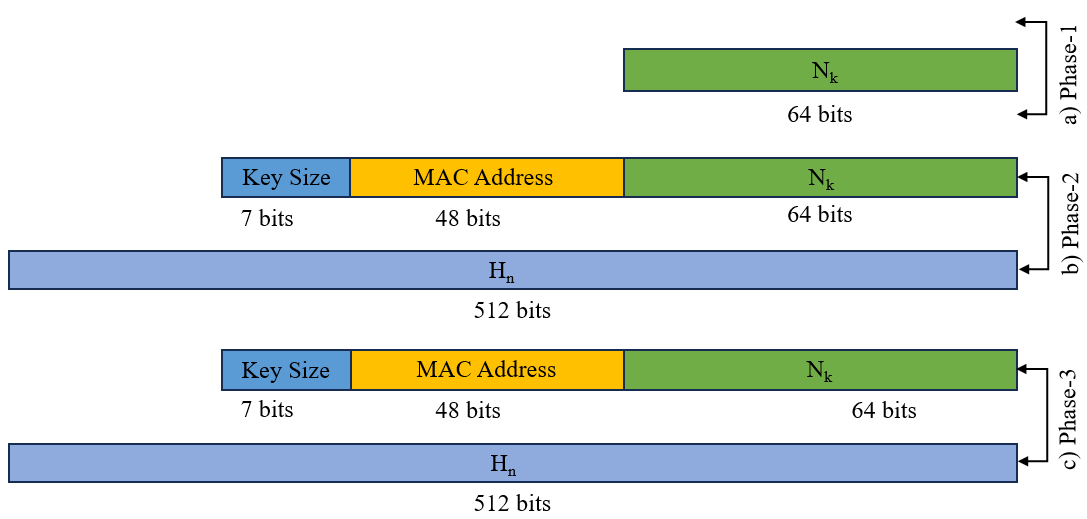}}
  \caption{Message flow size in each phase.}
  \label{fig}
\end{figure}

\subsection{Defense Techniques}
Though Blockchain offers amazing possibilities, some security challenges remain. These challenges are tackled head-on through the implementation of targeted solutions against the four most prominent attacks: 51\% attacks (adaptive hash rate increase), phishing attacks (domain verification and email checks), routing attacks (data encryption and robust routing validation), and Sybil attacks (transaction filtering and rigorous node validation)\cite{aggarwal2021attacks}. The following discussion will help to understand how these defense mechanisms improve the framework's strength briefly.

51\% Attack: A dynamic hash rate escalation mechanism is implemented to fortify the blockchain network against 51\% attacks. This mechanism strategically augments the network's hash rate after each block generation, simulating a progressive increase in difficulty. This simulated difficulty hinders attackers from seeking majority control, effectively thwarting their efforts. Empirical testing demonstrates the efficacy of this approach, as the modified prevention code consistently yielded the lowest attack success rates. This makes it much harder for attackers to take over the network by controlling its computing power, making it stronger against these attacks.

Phishing Attack: To mitigate phishing attacks within the model's deployed environment, additional measures are integrated. A whitelist of email sender domains is checked against incoming messages, with their existence further corroborated through DNS MX record checks. This multi-layered approach effectively detects spoofed addresses, a hallmark of phishing attempts. Even though clever tricks might pop up, our security features still make phishing scams way less likely, keeping the whole system much safer.

Routing Attack: Data encryption protocol has been used to secure transactions for each block. A strong password has been provided for the blocks to authentication. It helps us to make the model strong enough to defend the most routing attacks. This solution defines a function \textit{is\textunderscore valid\textunderscore destination()} that checks if a destination IP is a valid IPv4 or IPv6 address. It also defines a function \textit{route\textunderscore packet()} that takes the source IP and destination IP as arguments and checks if the destination IP is within the valid network range. If the destination IP is valid and within the network range, the packet is allowed to be routed. Otherwise, the packet is dropped. This mechanism usage demonstrates how to call the \textit{route\textunderscore packet()} function to route packets between different devices. The first packet is allowed to be routed because the destination IP is within the valid network range. The second packet is dropped because the destination IP is outside the network range.

Sybil Attack: To defend against Sybil attacks, a multi-layered defense strategy has been implemented. First, the algorithm exclusively operates on verified blocks, isolating potentially compromised ones. Second, a robust constructed a robust node validation system featuring a Node class with unique IDs, IP addresses, and reputation scores. This rigorous process, coupled with IP address validation, effectively thwarts the infiltration of malicious nodes, bolstering the network's resilience against Sybil attacks.

Blockchain security is an ongoing process that requires continuous vigilance and adaptation to evolving threats. By implementing robust security measures, educating users, and fostering a collaborative security culture, the blockchain ecosystem can mature and become more resilient against

\section{Conclusion and Future Research}
This paper proposes a novel SoftPUF-based blockchain framework for secure authentication. By leveraging SoftPUF's software-based key generation and integrating it with a blockchain network, the framework overcomes the hardware limitations of traditional PUFs and enables secure and scalable authentication for a wider range of devices. The framework facilitates secure communication, fast transaction times, and incorporates robust defense mechanisms, paving the way for secure and efficient authentication in diverse blockchain applications. Building upon this framework, future research can explore several promising directions. Optimizing the machine learning model within the blockchain context, for instance through reinforcement learning, could further enhance performance and accuracy. Additionally, investigating the integration of the framework with different blockchain platforms can assess its adaptability. Evaluating the framework's effectiveness in specific application scenarios, such as cloud-based systems or IoT deployments, will provide valuable insights into its real-world applicability. Finally, continued research on potential vulnerabilities and attack vectors specific to the framework's interaction within the blockchain network remains crucial for ensuring long-term security and robustness.

\bibliographystyle{IEEEtran}
\bibliography{bibfile}

\begin{thebibliography}{10}
\providecommand{\url}[1]{#1}
\csname url@samestyle\endcsname
\providecommand{\newblock}{\relax}
\providecommand{\bibinfo}[2]{#2}
\providecommand{\BIBentrySTDinterwordspacing}{\spaceskip=0pt\relax}
\providecommand{\BIBentryALTinterwordstretchfactor}{4}
\providecommand{\BIBentryALTinterwordspacing}{\spaceskip=\fontdimen2\font plus
\BIBentryALTinterwordstretchfactor\fontdimen3\font minus \fontdimen4\font\relax}
\providecommand{\BIBforeignlanguage}[2]{{%
\expandafter\ifx\csname l@#1\endcsname\relax
\typeout{** WARNING: IEEEtran.bst: No hyphenation pattern has been}%
\typeout{** loaded for the language `#1'. Using the pattern for}%
\typeout{** the default language instead.}%
\else
\language=\csname l@#1\endcsname
\fi
#2}}
\providecommand{\BIBdecl}{\relax}
\BIBdecl

\bibitem{bathalapalli2022pufchain}
V.~K. Bathalapalli, S.~P. Mohanty, E.~Kougianos, B.~K. Baniya, and B.~Rout, ``Pufchain 3.0: Hardware-assisted distributed ledger for robust authentication in the internet of medical things,'' in \emph{IFIP International Internet of Things Conference}.\hskip 1em plus 0.5em minus 0.4em\relax Springer, 2022, pp. 23--40.

\bibitem{shamsoshoara2020survey}
A.~Shamsoshoara, A.~Korenda, F.~Afghah, and S.~Zeadally, ``A survey on physical unclonable function (puf)-based security solutions for internet of things,'' \emph{Computer Networks}, vol. 183, p. 107593, 2020.

\bibitem{lee2020novel}
J.~Lee, S.~Choi, D.~Kim, Y.~Choi, and W.~Sun, ``A novel hardware security architecture for iot device: Pd-crp (puf database and challenge--response pair) bloom filter on memristor-based puf,'' \emph{Applied Sciences}, vol.~10, no.~19, p. 6692, 2020.

\bibitem{sadhu2022machine}
P.~K. Sadhu, A.~Baul, V.~P. Yanambaka, and A.~Abdelgawad, ``Machine learning and puf based authentication framework for internet of medical things,'' in \emph{2022 International Conference on Microelectronics (ICM)}.\hskip 1em plus 0.5em minus 0.4em\relax IEEE, 2022, pp. 160--163.

\bibitem{aarella2023fortified}
S.~G. Aarella, S.~P. Mohanty, E.~Kougianos, and D.~Puthal, ``Fortified-edge: Secure puf certificate authentication mechanism for edge data centers in collaborative edge computing,'' in \emph{Proceedings of the Great Lakes Symposium on VLSI 2023}, 2023, pp. 249--254.

\bibitem{gope2021scalable}
P.~Gope, O.~Millwood, and B.~Sikdar, ``A scalable protocol level approach to prevent machine learning attacks on physically unclonable function based authentication mechanisms for internet of medical things,'' \emph{IEEE Transactions on Industrial Informatics}, vol.~18, no.~3, pp. 1971--1980, 2021.

\bibitem{li2022blockchain}
D.~Li, R.~Chen, D.~Liu, Y.~Song, Y.~Ren, Z.~Guan, Y.~Sun, and J.~Liu, ``Blockchain-based authentication for iiot devices with puf,'' \emph{Journal of Systems Architecture}, vol. 130, p. 102638, 2022.

\bibitem{almuqren2023blockchain}
L.~Almuqren, K.~Mahmood, S.~S. Aljameel, A.~S. Salama, G.~P. Mohammed, and A.~A. Alneil, ``Blockchain assisted secure smart home network using gradient based optimizer with hybrid deep learning model,'' \emph{IEEE Access}, 2023.

\bibitem{jia2022blockchain}
X.~Jia, M.~Luo, H.~Wang, J.~Shen, and D.~He, ``A blockchain-assisted privacy-aware authentication scheme for internet of medical things,'' \emph{IEEE Internet of Things Journal}, vol.~9, no.~21, pp. 21\,838--21\,850, 2022.

\bibitem{patil2020blockchain}
A.~S. Patil, R.~Hamza, H.~Yan, A.~Hassan, and J.~Li, ``Blockchain-puf-based secure authentication protocol for internet of things,'' in \emph{Algorithms and Architectures for Parallel Processing: 19th International Conference, ICA3PP 2019, Melbourne, VIC, Australia, December 9--11, 2019, Proceedings, Part II 19}.\hskip 1em plus 0.5em minus 0.4em\relax Springer, 2020, pp. 331--338.

\bibitem{mahmood2023blockchain}
K.~Mahmood, S.~Shamshad, M.~A. Saleem, R.~Kharel, A.~K. Das, S.~Shetty, and J.~J. Rodrigues, ``Blockchain and puf-based secure key establishment protocol for cross-domain digital twins in industrial internet of things architecture,'' \emph{Journal of Advanced Research}, 2023.

\bibitem{felicetti2023deep}
C.~Felicetti, A.~Guzzo, G.~Manco, F.~Pasqua, E.~Ritacco, A.~Rullo, and D.~Sacc{\`a}, ``Deep learning/puf-based item identification for supply chain management in a distributed ledger framework,'' in \emph{2023 Fifth International Conference on Blockchain Computing and Applications (BCCA)}.\hskip 1em plus 0.5em minus 0.4em\relax IEEE, 2023, pp. 28--35.

\bibitem{mursi2021hybrid}
K.~T. Mursi, B.~Thapaliya, and Y.~Zhuang, ``A hybrid-optimizer-enhanced neural network method for the security vulnerability study of multiplexer arbiter pufs,'' in \emph{Journal of Physics: Conference Series}, vol. 1729, no.~1.\hskip 1em plus 0.5em minus 0.4em\relax IOP Publishing, 2021, p. 012010.

\bibitem{aggarwal2021attacks}
S.~Aggarwal and N.~Kumar, ``Attacks on blockchain,'' in \emph{Advances in computers}.\hskip 1em plus 0.5em minus 0.4em\relax Elsevier, 2021, vol. 121, pp. 399--410.

\end{thebibliography}

\end{document}